\documentstyle[amstex,amssymb,preprint,aps]{revtex}

\begin{document}
\title{THEORY OF COMPLEX SCATTERING LENGTHS}
\author{M.S. Hussein}
\address{Instituto de F\'\i sica, Universidade de S\~ao Paulo, \\
C.P. 66318, S\~{a}o Paulo, 05315-970, Brazil}
\date{\today}
\maketitle

\begin{abstract}
We derive a generalized Low equation for the T-matrix appropriate for
complex atom-molecule interaction. The properties of this new equation at
very low energies are studied and the complex scattering length and
effective range are derived.
\end{abstract}

\newpage

The recent realization of Bose-Einstein condensation (BEC) of ultracold
atoms with the accompanying upsurge of theoretical activities have rekindled
interest in low energy collisions of atoms and molecules. The subsequent
proposals for the creation of ultracold molecular [1-5] and hybrid
atomic-molecular BEC [6,7] intensified the above mentioned interest \ Of
particular importance in the above recent developments is the idea of decay
of the condensates. In a series of papers, Dalgarno and collaborators [8-12]
have looked into the idea of using a complex scattering length to represent
the low-energy atom-molecule scattering. Implicit in this is the
multichannel nature of the collision process: an atom hits a vibrationally
excited molecule at extremely low energies. The open inelastic channels are
those where the molecule is excited into lower vibrational states. In this
sense one has a depletion of the elastic channel. In Ref. [8], the quenching
of $H_{2}$ molecules in collisions with $H$ was considered. It was found
that, the inelastic cross-sections and the corresponding depletion rate
coefficients were very large for high vibrational levels of $H_{2}$.

In the above studies, the following form of low-energy S-wave scattering
amplitude is used

\begin{equation}
f\left( k\right) =\frac{1}{g\left( k^{2}\right) -i\text{ }k}\text{ ,}
\end{equation}

\noindent where k is the wave number related to the center of mass energy of
the colliding partners, $E$, by $\frac{\hslash k^{2}}{2\mu }=E$, with $\mu $
being the reduced mass of the system. The function $g\left( k^{2}\right) $
is even in $k$ and is given by the effective range formula.

\begin{equation}
g\left( k^{2}\right) =-\frac{1}{a}+\frac{1}{2}r_{o}k^{2}\text{ ,}
\end{equation}

\noindent where a is the scattering length and $r_{\circ }$ the effective
range, both directly related to the interaction. When applied to
atom-molecule scattering at very low energies, with the molecules suffering
inelastic transitions to lower vibrational states, the scattering length a
is taken to be complex, $a=\alpha -i\beta $, with $\beta $ related to the
total inelastic cross-section.

The question we raise here is the validity of Eq. (1) with a and eventually $%
r_{\circ }$ taken as complex in the case of the elastic scattering with
strong coupling to inelastic channels. Of course, an equivalent one-channel
description of the elastic scattering can be formulated with the
introduction of an appropriate complex optical potential as described by
Feshbach [13]. It is therefore legitimate to inquire about the validity of
Eq. $\left( 1\right) $, originally obtained for real potential, if a complex
interaction is used [14]. The general structure of the low energy scattering
amplitude is also of potentially fundamental importance to very low energy
matter interferometry. This method for the obtention of $f$ for
molecule-molecule scattering has been quite successful at room\ temperatures
[15,16]. Extension to very low temperatures of this method seems natural and
would welcome studies of the type reported here.

For the above purpose, it is useful to summarize the elegant derivation of
Eq. $\left( 1\right) $ given by Weinberg [17]. If we denote the interaction
by $V$ and the free Green's function by $G_{o}^{\left( +\right) }\left(
E\right) =\left( E-H_{o}+i\varepsilon \right) ^{-1}$, then the T-matrix
given by the Lippmann-Schwinger equation $T^{\left( +\right)
}=V+VG_{o}^{\left( +\right) }T^{\left( +\right) }$, can be written\ as $%
T^{\left( +\right) }=V+VG^{\left( +\right) }V$, with the full Green's
function $G^{\left( +\right) }=\left( E+i\varepsilon -H_{o}-V\right) ^{-1}$.
Using the spectral expansion of $G^{\left( +\right) }$, with the complete
set of bound and scattering states $\left\{ \left| B\right\rangle ,\left|
\Psi ^{\left( +\right) }\right\rangle \right\} $, we obtain the Low-equation

\begin{equation}
\left\langle \vec{k}\prime \left| T^{\left( +\right) }\left( E\right)
\right| \vec{k}\right\rangle =\left\langle \vec{k}\prime \left| V\right| 
\vec{k}\right\rangle +%
\mathop{\displaystyle\sum}%
_{B}\frac{\left\langle \vec{k}\prime \left| V\right| B\right\rangle
\left\langle B\left| V\right| \vec{k}\right\rangle }{E+E_{B}}+%
\displaystyle\int %
d\vec{k}^{\prime \prime }\frac{T_{\vec{k}\prime k^{\prime \prime }}^{\left(
+\right) }\left( E_{k\prime \prime }\right) \left( T_{\vec{k}\prime \prime
k}^{\left( +\right) }\left( E_{k\prime \prime }\right) \right) ^{\ast }}{%
E-E_{k^{\prime \prime }}+i\varepsilon }\text{ .}
\end{equation}

At very low energies relevant for BEC, we seek a solution $T_{\vec{k}\prime
k}\left( E\right) \equiv T\left( E\right) $ and writing $\left\langle \vec{k}%
\prime \left| V\right| \vec{k}\right\rangle =$ $\bar{V}$ we have

\begin{equation}
t^{\left( +\right) }\left( E\right) =\bar{V}+\sum\limits_{B}\frac{\left|
g_{B}\right| ^{2}}{E+E_{B}}+\int d\vec{k}^{\prime \prime }\frac{\left|
t^{\left( +\right) }\left( E_{k^{\prime \prime }}\right) \right| ^{2}}{%
E-E_{k^{\prime \prime }}+i\varepsilon }\text{ .}
\end{equation}

Calculating now $t^{\left( +\right) }\left( E\right) ^{-1}-t^{\left(
-\right) }\left( E\right) ^{-1}$, we find

\begin{equation}
t^{\left( +\right) }\left( E\right) ^{-1}-t^{\left( -\right) }\left(
E\right) ^{-1}=\frac{t^{\left( -\right) }\left( E\right) -t^{\left( +\right)
}\left( E\right) }{t^{\left( -\right) }\left( E\right) \text{ }t^{\left(
+\right) }\left( E\right) }\text{ .}
\end{equation}

Since $t^{\left( -\right) }\left( E\right) =T\left( E-i\varepsilon \right)
=\left( T^{\left( +\right) }\left( E+i\varepsilon \right) \right) ^{\ast }$,
if $V$ is real, we have $t^{\left( +\right) }\left( E\right) ^{-1}-t^{\left(
-\right) }\left( E\right) ^{-1}=-2ik$ $2\pi \frac{2\mu }{\hbar ^{2}}$ which
is just the discontinuity across the positive energy cut in the complex
energy plane. Besides the poles in $t$, $\left( \text{zeros in }\left(
t\right) ^{-1}\right) $, the only other terms in $\left( t^{\left( +\right)
}\right) ^{-1}$ are entire functions of $W\equiv E+i\varepsilon $.
Accordingly, with the identification$\ f=-\frac{1}{2\pi }\frac{2\mu }{\hbar
^{2}}t$, Eq. (1) follows.

We turn next to a complex interaction $V\neq V^{\dagger }$. The completeness
relation now reads $\sum\limits_{B}\left| B\rangle \right. \langle B|+\int d%
\vec{k}\prime \prime \left| \Psi _{\vec{k}\prime \prime }^{\left( +\right)
}\rangle \text{ }\langle \tilde{\Psi}_{k\prime \prime }^{\left( +\right)
}\right| $ where $\left| \tilde{\Psi}_{\vec{k}\prime \prime }^{\left(
+\right) }\right. \rangle $ is the dual scattering state which is a solution
of the Schr\"{o}dinger equation with $V$ replaced by $V^{\dagger }$ [18,19].
Another form of the completeness relation may also be used, $%
\sum\limits_{B}\left| B\rangle \right. \langle B|+\int d\vec{k}\prime \prime
\left| \tilde{\Psi}_{\vec{k}\prime \prime }^{\left( -\right) }\rangle \text{ 
}\langle \Psi _{k\prime \prime }^{\left( -\right) }\right| $,with $\left|
\Psi _{\vec{k}\prime \prime }^{\left( -\right) }\right\rangle $ being the
physical scattering state with incoming wave boundary condition $\left(
V^{\dagger }\text{,}-i\varepsilon \right) $ and $\left| \tilde{\Psi}_{\vec{k}%
\prime \prime }^{\left( -\right) }\right\rangle $ its corresponding dual
state $\left( V\text{,}-i\varepsilon \right) $.\ Thus, the full Green's
function now has the spectral form

\begin{equation}
G^{\left( +\right) }\left( E\right) =\sum_{B}\frac{\left| B\rangle \right.
\langle B|}{E+E_{B}}+\int d\vec{k}\prime \prime \frac{\left| \Psi _{\vec{k}%
}^{\left( +\right) }\rangle \text{ }\langle \tilde{\Psi}_{\vec{k}\prime
\prime }^{\left( +\right) }\right| }{E-E_{k\prime \prime }+i\varepsilon }%
\text{ .}
\end{equation}

Accordingly, Eq. (3) now reads

\begin{equation}
\left\langle \vec{k}\prime \left| T\right| \vec{k}\right\rangle
=\left\langle \vec{k}\prime \left| V\right| \vec{k}\right\rangle +\sum_{B}%
\frac{\left\langle \vec{k}\prime \left| V\right| B\right\rangle \left\langle
B\left| V\right| \vec{k}\right\rangle }{E+E_{B}+i\varepsilon }+\int d\vec{k}%
\prime \prime \frac{\left\langle \vec{k}\prime \left| V\right| \Psi _{\vec{k}%
\prime \prime }^{\left( +\right) }\right\rangle \left\langle \tilde{\Psi}_{%
\vec{k}\prime \prime }^{\left( +\right) }\left| V\right| \vec{k}%
\right\rangle }{E-E_{k\prime \prime }+i\varepsilon }\text{ .}
\end{equation}

It is clear that the Low equation, Eq. (3), is not valid anymore. However,
as we show below Eq. (1) is still valid, with the appropriate generalization
of the real function $g\left( k^{2}\right) $ to a complex one [16]. To see
this we analyze the matrix element $\left\langle \tilde{\Psi}_{\vec{k}\prime
\prime }^{\left( +\right) }\left| V\right| \vec{k}\right\rangle $. From the $%
L-S$ equation for $\left\langle \tilde{\Psi}_{\vec{k}\prime \prime }^{\left(
+\right) }\right| $,

\begin{equation}
\left\langle \tilde{\Psi}_{\vec{k}\prime \prime }^{\left( +\right) }\right|
=\left\langle \vec{k}\prime \right| +\left\langle \vec{k}\prime \right| V%
\frac{1}{E_{k\prime \prime }-H_{\circ }-V-i\varepsilon }\equiv \left\langle 
\vec{k}\prime \right| \left[ 1+VG^{\left( -\right) }\left( E_{k\prime \prime
}\right) \right] \text{ .}
\end{equation}

\noindent Thus $\left\langle \tilde{\Psi}_{\vec{k}\prime \prime }^{\left(
+\right) }\left| V\right| \vec{k}\right\rangle =\left\langle \vec{k}\prime
\left| \tilde{T}\left( E_{k\prime \prime }-i\varepsilon \right) \right| \vec{%
k}\right\rangle $, where the unphysical T-matrix $\tilde{T}$ is given by

\begin{equation}
\tilde{T}=V+VG^{\left( -\right) }V\text{ .}
\end{equation}

Accordingly the T-matrix equation, Eq (7), may be written as

\begin{gather}
\left\langle \vec{k}\prime \left| T\left( E\right) \right| \vec{k}%
\right\rangle =\left\langle \vec{k}\left| V\right| \vec{k}\right\rangle
+\sum_{B}\frac{\left\langle \vec{k}\prime \left| V\right| B\right\rangle
\left\langle B\left| V\right| \vec{k}\right\rangle }{E+E_{B}+i\varepsilon } 
\nonumber \\
+\int d\vec{k}\prime \prime \frac{\left\langle \vec{k}\prime \left| T\left(
E\prime \prime \right) \right| \vec{k}\prime \prime \right\rangle
\left\langle \vec{k}\prime \prime \left| \tilde{T}\left( E\prime \prime
\right) \right| \vec{k}\right\rangle }{E-E\prime \prime +i\varepsilon }\text{
.}
\end{gather}

A similar equation holds for $\left\langle \vec{k}\prime \left| \tilde{T}%
\left( E\right) \right| \vec{k}\right\rangle $with $i\varepsilon $ replaced
by $-i\varepsilon $. It is interesting at this point to show the relation
between the physical T-matrix element $\left\langle \vec{k}\prime \left|
T\left( E\right) \right| \vec{k}\right\rangle $ and $\left\langle \vec{k}%
\prime \left| \tilde{T}\left( E\right) \right| \vec{k}\right\rangle $. This
can be done easily following operator manipulations of [18], and using the
relation $\left\langle \tilde{\Psi}_{\vec{k}\prime \prime }^{\left( +\right)
}\right| =\left\langle \Psi _{\vec{k}\prime \prime }^{\left( +\right)
}\right| +\left\langle \Psi _{\vec{k}\prime \prime }^{\left( +\right)
}\right| \left( V-V^{\dagger }\right) G^{\left( -\right) }\left( E_{k\prime
\prime }\right) $, Eq. (8),

\begin{equation}
\left\langle \vec{k}\prime \left| \tilde{T}\left( E\right) \right| \vec{k}%
\right\rangle =\left\langle \vec{k}\prime \left| T\left( E\right) \right| 
\vec{k}\right\rangle ^{\ast }+\int d\vec{k}\prime \prime \left\langle \Psi _{%
\vec{k}\prime }^{\left( +\right) }\left| \left( V-V^{^{\dagger }}\right)
\right| \Psi _{\vec{k}\prime \prime }^{\left( +\right) }\right\rangle S_{%
\vec{k}\prime \prime \vec{k}}^{-1}\text{ ,}
\end{equation}

\noindent where $S^{-1}$ is the inverse S-matrix in the elastic channel, $S_{%
\vec{k}\prime \prime \vec{k}}^{-1}=\left\langle \tilde{\Psi}_{\vec{k}\prime
\prime }^{\left( +\right) }\right. \left| \tilde{\Psi}_{\vec{k}}^{\left(
-\right) }\right\rangle $, and the diagonal part of the matrix element $%
\left\langle \Psi _{\vec{k}\prime }^{\left( +\right) }\left| \left(
V-V^{\dagger }\right) \right| \Psi _{\vec{k}\prime \prime }^{\left( +\right)
}\right\rangle $is directly related to the total inelastic scattering
cross-section, $\sigma _{in}$, viz [18]

\begin{equation}
\left\langle \Psi _{\vec{k}}^{\left( +\right) }\left| \left( V-V^{\dagger
}\right) \right| \Psi _{\vec{k}}^{\left( +\right) }\right\rangle =-2i\frac{E%
}{k}\sigma _{in}\left( E\right) \text{ .}
\end{equation}

Eq. (11) explicitly exhibits the connection between $\tilde{T}$ and $T$
through the absorptive part of the effective interaction.

Now we seek the low energy solution $\left\langle \vec{k}\prime \left|
T\right| \vec{k}\right\rangle \equiv t_{+}\left( E\right) $ and $%
\left\langle \vec{k}\prime \left| \tilde{T}\right| \vec{k}\right\rangle
\equiv t_{-}\left( E\right) $ and following the same steps as Weinberg's
[17], we find immediately, from Eq. (10), with $f_{\pm }=-\frac{1}{2\pi }%
\frac{2\mu }{\hbar ^{2}}t_{\pm }$,

\begin{equation}
f_{+}^{-1}=g_{c}\left( k^{2}\right) -ik\text{ ;}
\end{equation}

\begin{equation}
f_{-}^{-1}=g_{c}\left( k^{2}\right) +ik\text{ ,}
\end{equation}

\noindent where $g_{c}\left( k^{2}\right) $ is the complex generalization of 
$g\left( k^{2}\right) $ of Eq. (1).

We turn now to the connection between $g_{c}\left( k^{2}\right) $ and the
low-energy observables. This is most conveniently accomplished by employing
the generalized optical theorem

\begin{equation}
\frac{4\pi }{k}%
\mathop{\rm Im}%
f_{+}=\sigma _{el}+\sigma _{in}\text{ ,}
\end{equation}

\noindent where $\sigma _{el}$ is the total elastic scattering cross section 
$4\pi \left| f_{+}\right| ^{2}$ and $\sigma _{in}$ the total inelastic
cross-section.

Using (12), we find

\begin{equation}
\frac{-%
\mathop{\rm Im}%
\text{ }g_{c}\left( k^{2}\right) }{\left( 
\mathop{\rm Re}%
\text{ }g_{c}\left( k^{2}\right) \right) ^{2}+\left( 
\mathop{\rm Im}%
\text{ }g_{c}\left( k^{2}\right) -k\right) ^{2}}=\frac{k}{4\pi }\sigma _{in}%
\text{ .}
\end{equation}

At $k=0$, $g_{c}\left( 0\right) =-\frac{1}{a}$, where a is the complex
scattering length written as [8] $\alpha -i\beta $. Thus the imaginary part
of a, $\beta $, is found to be

\begin{equation}
\beta =\frac{\left( k\text{ }\sigma _{in}\right) _{k=0}}{4\pi }\text{ ,}
\end{equation}

\noindent an expression also derived in Ref. [8]. Eq. (16) clearly implies
that $\sigma _{in}$ should go as $k^{-1}$ as $k$ is lowered, in accordance
with Wigner's law.

We go a bit beyond Refs. [8-12] and derive a relation between $\beta $ and
the imaginary part of the effective potential. Since $\sigma _{in}$ is given
by (for S-wave scattering), Eq. (12)

\begin{equation}
\sigma _{in}=\frac{4\pi }{kE}\int\limits_{0}^{\infty }\left| u\left(
r\right) \right| ^{2}\left| 
\mathop{\rm Im}%
V\text{ }\left( r\right) \right| \text{ }dr\text{ ,}
\end{equation}

\noindent where $u\left( r\right) $ is the S-wave elastic radial wave
function, we find

\begin{equation}
\beta =\left[ \frac{1}{E}\int\limits_{0}^{\infty }\left| u\left( r\right)
\right| ^{2}\left| 
\mathop{\rm Im}%
V\text{ }\left( r\right) \right| \text{ }dr\right] _{E\rightarrow 0}\text{ .}
\end{equation}

An equation for the complex effective range, $r_{\circ }$, can also be
easily derived. Equation (19) is the principle result of this work. It
summarizes the following:

\begin{enumerate}
\item[1)]  The coupled-channels calculation aimed to describe the molecular
quenching can be recast as an effective one-channel calculation with a
complex interaction whose imaginary part account for flux loss.

\item[2)]  The low-energy behaviour of the scattering amplitude with the 
\underline{complex} interaction alluded to above can be conveniently
parametrized in terms of \underline{complex} scattering length and effective
range.
\end{enumerate}

The message this work conveys is the potential usefulness of constructing
the effective complex (optical) interaction for the scattering of
ro-vibrational molecules from atoms at low energies. The calculation of a
and $r_{\circ }$ from knowledge of this potential can be done in a direct
and unambiguous way.\vspace{1.1cm}

{\bf Acknowledgement}{\Huge \bigskip }

Part of this work was done while the author was visiting ITAMP-Harvard. He
wishes to thank Prof. Kate Kirby and Dr. H. Sadeghpour for hospitality. He
also thanks Drs. N. Balakrishnan and V. Kharchenko for useful discussion.

Partial support form the ITAMP-NSF grant and from FAPESP and CNPq is
acknowledged.

\end{document}